\documentclass[11pt]{article}
\usepackage{moriond}

\def\Journal#1#2#3#4{{#1} {\bf #2}, #3 (#4)}

\def\PRL{\em Phys. Rev. Lett.}
\def\PRC{{\em Phys. Rev.} C}
\def\PRD{{\em Phys. Rev.} D}
\def\ASTR{\em Asto. Part. Physics}

\def\EPJ{{\em Eur. Physics. J.} C}

\def\be{\begin{equation}}
\def\ee{\end{equation}}
\def\bea{\begin{eqnarray}}
\def\eea{\end{eqnarray}}

\def\kevee{keV$_{\mathrm{ee}}$}
\def\kevnr{keV$_{\mathrm{nr}}$}
\def\leff{$\mathcal{L}_{eff}$}
\def\qy{$\mathcal{Q}_{y}$}

\begin{document}
\vspace*{4cm}
\title{Direct Dark Matter search with the XENON program}

\author{P. Beltrame (on behalf of the XENON Collaboration)}
\address{Dept. of Particle Physics \& Astrophysics, Weizmann Institute of Science,\\ Herzl Str. 234, 76100 Rehovot, Israel}

\maketitle

\abstracts{We present the most recent results from XENON100, the current phase of the XENON dark matter search program. XENON100 is a dual phase time-projection chamber operated at the Laboratori Nazionali del Gran Sasso (LNGS) whose ultra-low electromagnetic background, about $5 \times 10^{-3}$ events/($\mathrm{kg}\times\mathrm{day}\times\mathrm{keV}$), allowed to set the most stringent limit to date, excluding WIMP-nucleon spin-independent interaction down to cross-sections of $2 \times 10^{-45}$ cm$^2$ for a 55 GeV/c$^2$ mass at 90\% confidence level and $3.5 \times 10^{-40}$ for 45 GeV/c$^2$ in the spin-dependent interaction with neutrons. We also introduce the status and physics goal of XENON1T, the next phase of the program, which will be able to achieve sensitivity down to $2 \times 10^{-47}$ cm$^2$ for a WIMP of 50 GeV/c$^2$.}

\section{Dark Matter direct detection by XENON}

\begin{figure}
\begin{minipage}{0.5\linewidth}
\centerline{\includegraphics[angle=-90, width=1.\linewidth]{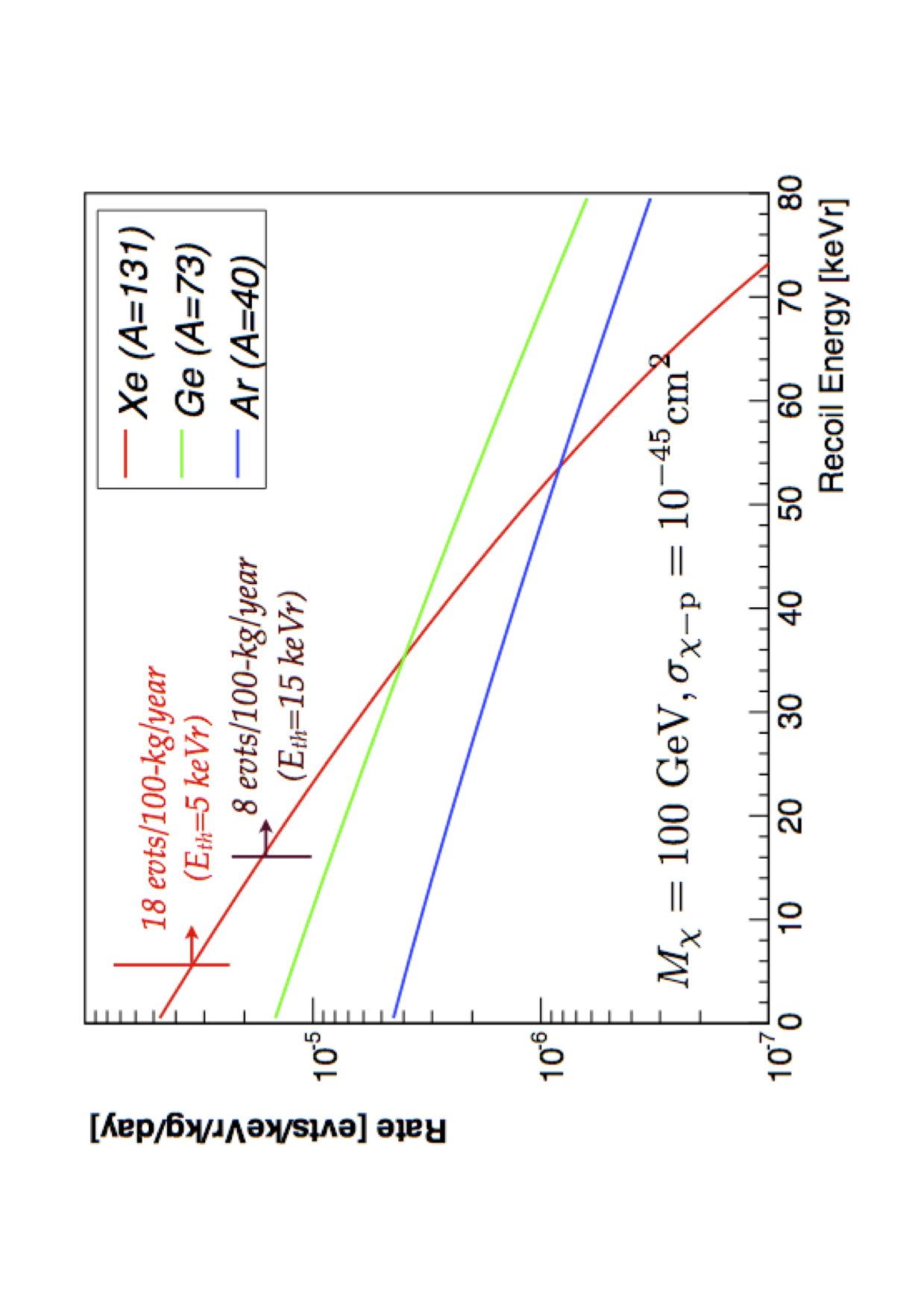}}
\end{minipage}
\begin{minipage}{0.5\linewidth}
\centerline{\includegraphics[width=1.\linewidth]{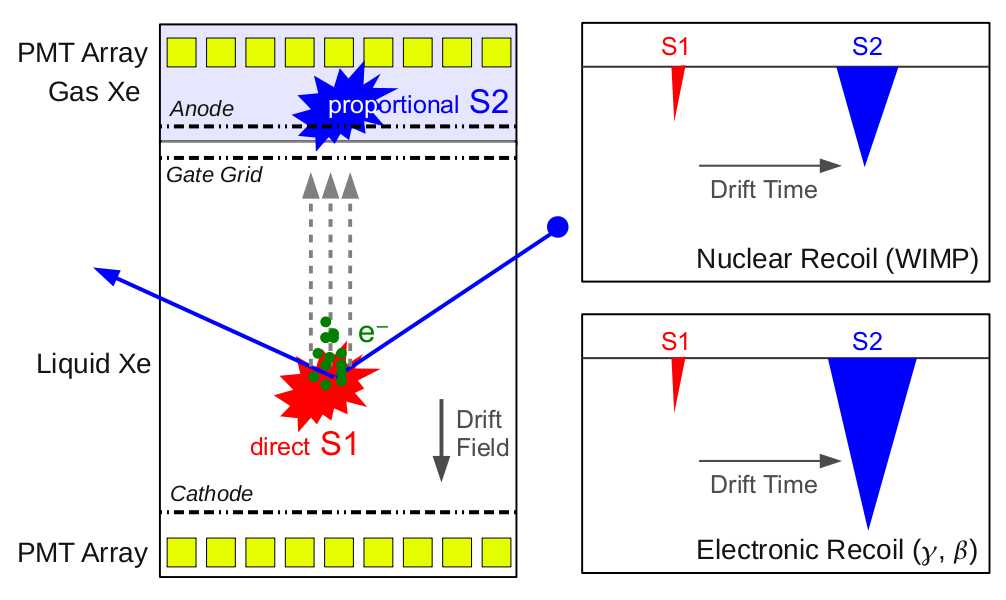}}
\end{minipage}
\caption{Left: Rate in events/($\mathrm{kg}\times\mathrm{day}\times\mathrm{keV}$) for a 100 GeV/c$^2$ WIMP assuming a cross section $\sigma = 10^{-45}$ cm$^2$. Right: Schematic of the particle interaction inside a double phase Xe TPC.}
\label{fig:xe100wimp}
\end{figure}

Astronomical and cosmological observations indicate that a large amount of the energy content of the Universe is made of dark matter. Particle candidates, under the generic name of Weakly Interacting Massive Particles (WIMPs), arise naturally in many theories beyond the Standard Model of particle physics.\cite{dm} On Fig.\ref{fig:xe100wimp} left it is shown the expected signal, in events/($\mathrm{kg}\times\mathrm{day}\times\mathrm{keV}$ (DRU) of a WIMP in different target material. The extremely low rate and the recoil spectrum which falls exponentially with energy and extends to a few tens of keV only requires large detectors built from radio-pure materials and low energy threshold. \\
The XENON100 experiment aims to detect galactic dark matter through the elastic nuclear scattering of WIMPs from Xe target nuclei in a cylindrical (30 cm height x 30 cm diameter) dual phase time-projection chamber (TPC), operated at the Laboratori Nazionali del Gran Sasso (LNGS).\cite{xe100instr} A liquid Xe (LXe) target mass of 62 kg, is optically separated from an additional 99 kg, instrumented as an active scintillator veto. Both TPC and veto are mounted in a double-walled stainless-steel cryostat, enclosed by a passive shield. The detector is equipped of 242 radio pure photomultiplier tubes (PMTs) placed on top and on the bottom of the TPC. \\
A particle interaction in the LXe target creates both excited and ionized atoms. De-excitation leads to a prompt scintillation signal ($S1$). Due to the presence of an electric drift field of 530 V/cm, a large fraction of the ionization electrons is drifted away from the interaction site and extracted from the liquid into the gas phase by a strong extraction field of $\sim$12 kV/cm, generating a light signal ($S2$) by proportional scintillation in the gas. 3-dimensional event vertex reconstruction is achieved using the drift time of the electrons to reconstruct the $z$ position and the hit pattern on the PMTs in the gas phase to reconstruct the ($x,\; y$) position. See schematic drawing on Fig.\ref{fig:xe100wimp} right. \\
The ratio $S2/S1$ is different for electronic recoil events from interactions with the atomic electrons (ERs, from $\gamma$ and $\beta$ backgrounds), and for interactions with the nucleus itself (NRs, nuclear recoils from WIMPs or neutrons), and is used to discriminate the signal against background. The NR energy scale, measured in \kevnr ~(nuclear recoil equivalent), depends on a quenching factors called \leff ~and \qy, for $S1$ and $S2$ respectively. The energy scale calibration for ER interactions in LXe TPCs in general is referred as the electron-equivalent energy scale (\kevee) and it is determined by employing $\gamma$-ray emission lines from standard calibration sources, $^{60}$Co and $^{232}$Th.

\section{2011/2012 science run and $\sigma_{\chi-N}$ physics analysis}

\begin{figure}
\begin{minipage}{0.5\linewidth}
\centerline{\includegraphics[height=5 cm, width=1.\linewidth]{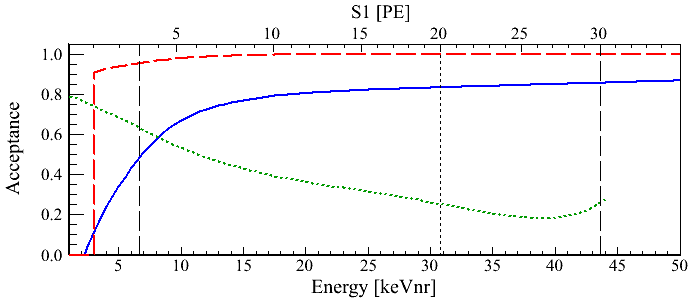}}
\end{minipage}
\begin{minipage}{0.5\linewidth}
\centerline{\includegraphics[width=1.\linewidth]{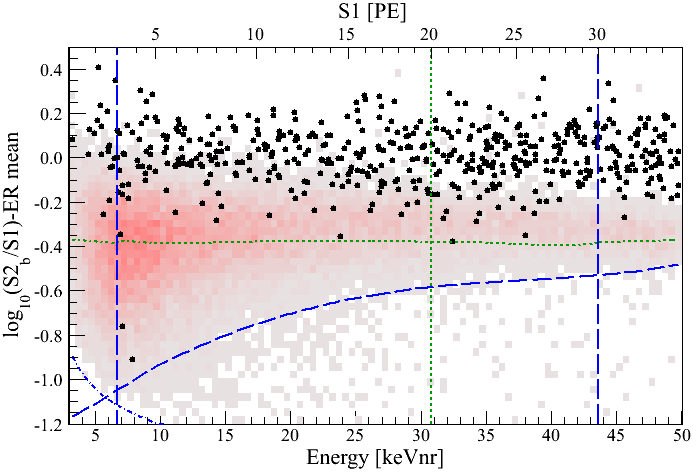}}
\end{minipage}
\caption{Left: Combined cut acceptance (solid blue). The S2 threshold cut $S2 > 150$ PE (dashed red). The acceptance is conservatively set to zero below 1 PE. NR acceptance is shown by the dotted green line. The lower analysis threshold is 6.6 \kevnr ~(3 PE) and extends to 43.3 \kevnr. Right: Event distribution in the discrimination parameter space $log_{10}(S2_b/S1)$ as observed after unblinding using all analysis cuts and a 34 kg fiducial volume (black points). An hard $S2_b/S1$ discrimination cut at 99.75\% ER rejection defines the benchmark WIMP search region from above (dotted green) only used as cross check. The histogram in red indicates the NR band from the neutron calibration.}
\label{fig:acceunblind}
\end{figure}

The latest XENON100 dark matter data was taken over a period of 13 months between February, 2011 and March, 2012, and, besides 3 interruptions due to equipment maintenance, was only interrupted by regular calibrations using blue LED light (for the PMT response), $^{137}$Cs source (for monitoring of the LXe purity), and $^{60}$Co and $^{232}$Th sources (for ER background calibration). To calibrate the response to NRs, data from an $^{241}$AmBe neutron source were taken just before the start and right after the end of the run. This results in final dark matter dataset of 224.6 live days. In order to avoid analysis bias, the dark matter data was blinded from $2 - 100$ photoelectrons (PE) in $S1$. \\
Compared to the previous data set,\cite{xe100run08} the new dark matter search is characterized by a considerably larger exposure and a substantial reduction of the intrinsic background from $^{85}$Kr (the natKr concentration was lowered down to $(19 \pm 4)$ ppt). In addition to that, an improved electronic noise conditions and a new hardware trigger allowed for a reduced $S2$  threshold leading to $> 99\%$ for $S2$ signals above 150 (PE), see red dashed line of Fig.\ref{fig:acceunblind} left.

All the analysis cuts efficiencies have been estimated on NR calibration data - with the exception of quality cuts which might have a time dependence due to changing detector conditions - and have been tested on the non-blind part of the science data or on the ER calibration data. The first class of cuts are basic data quality cuts. Since only single-scatter events are expected from WIMP interactions, the second class of cuts identifies such events using the number of $S1$ and $S2$ peaks in the waveform and the information from the active veto. Conditions on the size of the $S2$ and the requirement that at least two PMTs must observe an $S1$ peak ensure that only data above the threshold and well above the noise level are considered. The acceptances are shown in Fig.\ref{fig:acceunblind} left. Both the signal and the background-only hypothesis are tested by means of a Profile Likelihood (PL) statistical approach,\cite{xe100pl} while a maximum gap counting method with a predefined signal region - $6.6 - 30.5$ \kevnr ~($3 - 20$ PE) energy range and an upper 99.75\% ER rejection line (horizontal dashed green line on Fig.\ref{fig:acceunblind} left) - is used as a cross check. The fiducial volume used in this analysis contains 34 kg of LXe.\\ 
The NR background is determined from Monte Carlo simulations, with the input from precise measurements of all the detector components radioactivity and from muon-induced neutrons contamination. The total expectation from neutron background is $(0.17 + 0.12 - 0.07)$ events for the given exposure and NR acceptance in the benchmark region. ER background events originate from radioactivity of the detector components and from and activity of intrinsic radioactivity in the LXe target, such as $^{222}$Rn and $^{85}$Kr. To estimate the total ER background from all sources, the $^{60}$Co and $^{232}$Th calibration data are used, with $> 35$ times more statistics in the relevant energy range than in the dark matter data, leading to a total ER background expectation of $(0.79 \pm 0.16)$ in the benchmark region. \\
After unblinding, two events were observed in the benchmark WIMP search region, see Fig.\ref{fig:acceunblind} right, with energies of 7.1 \kevnr ~(3.3 PE) and 7.8 \kevnr ~(3.8 PE). The waveforms for both events are of high quality and their $S2/S1$ value is at the lower edge of the NR band from neutron calibration. There are no leakage events below 3 PE. The PL analysis yields a $p-$value of 5\% for all WIMP masses for the background-only hypothesis indicating that there is no excess due to a dark matter signal. The probability of the expected background to fluctuate in the benchmark to 2 events is 26.4\% and confirms this conclusion.

\subsection{Spin-independent cross section limit}
\label{subsec:run10si}

\begin{figure}
\centerline{\includegraphics[width=0.67\linewidth]{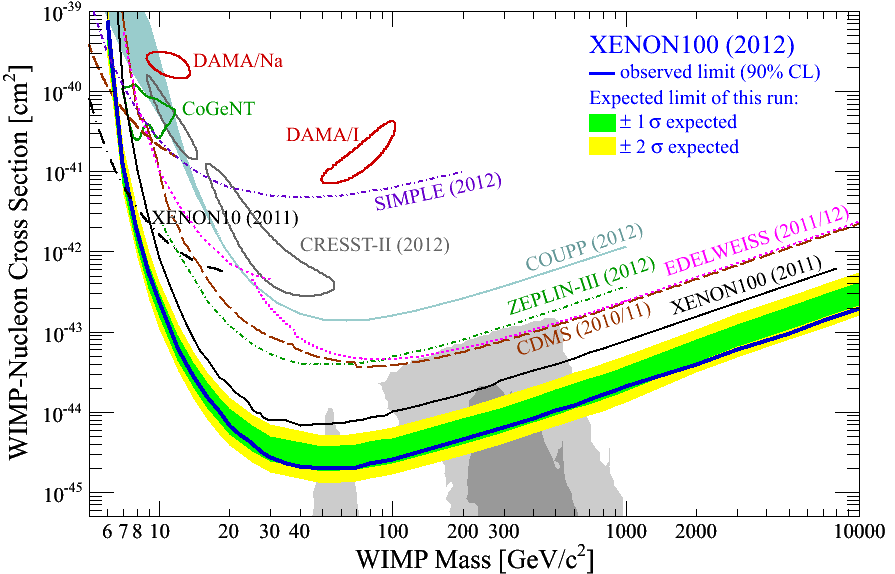}}
\caption{New result on spin-independent WIMP-nucleon scattering from XENON100.}
\label{fig:run10si}
\end{figure}

A 90\% confidence level exclusion limit for spin-independent WIMP-nucleon cross sections $\sigma_{\chi}$ is calculated. Standard assumptions on isothermal WIMP halo with a local density of $\rho_{\chi} = 0.3$ GeV/cm$^3$, a local circular velocity of $v_0 = 220$ km/s and a Galactic escape velocity of $v_{esc} = 544$ km/s have been made. Systematic uncertainties in the energy scale as described by the \leff ~parametrization and in the background expectation are profiled out. Poisson fluctuations in the number of PEs dominate the $S1$ energy resolution and are also taken into account. The expected sensitivity of this dataset in absence of any signal is shown by the green/yellow (1$\sigma$/2$\sigma$) band in Fig.\ref{fig:run10si}. The new limit is represented by the thick blue line, setting the best limit to date, excluding spin-independent WIMP-N cross-sections down to $2 \times 10^{-45}$ cm$^2$ for a 55 GeV/c$^2$ mass at 90\% confidence level.\cite{xe100run10si}

\subsection{Spin-dependent cross section limits}

\begin{figure}
\begin{minipage}{0.5\linewidth}
\centerline{\includegraphics[width=1.\linewidth]{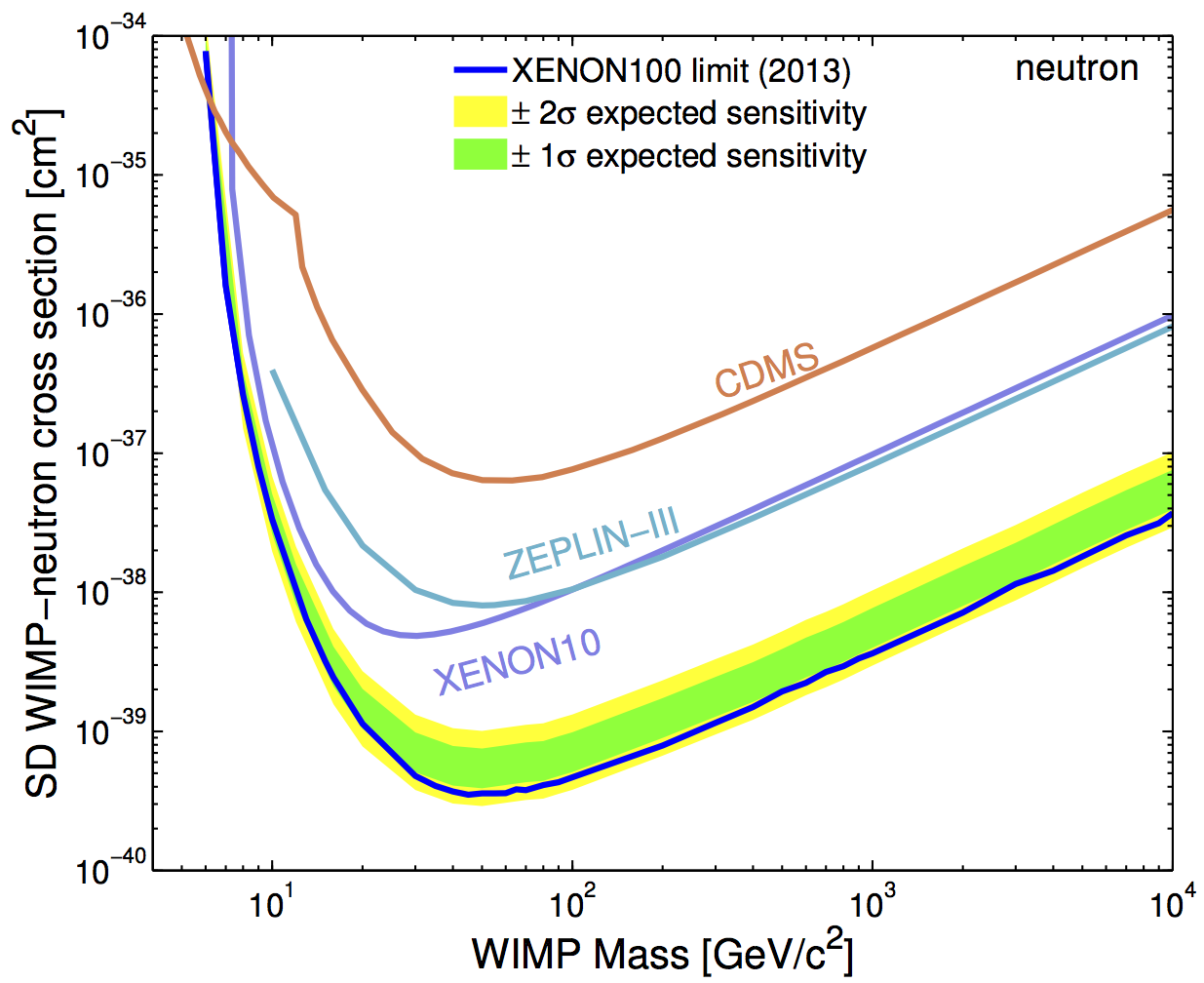}}
\end{minipage}
\begin{minipage}{0.5\linewidth}
\centerline{\includegraphics[width=1.\linewidth]{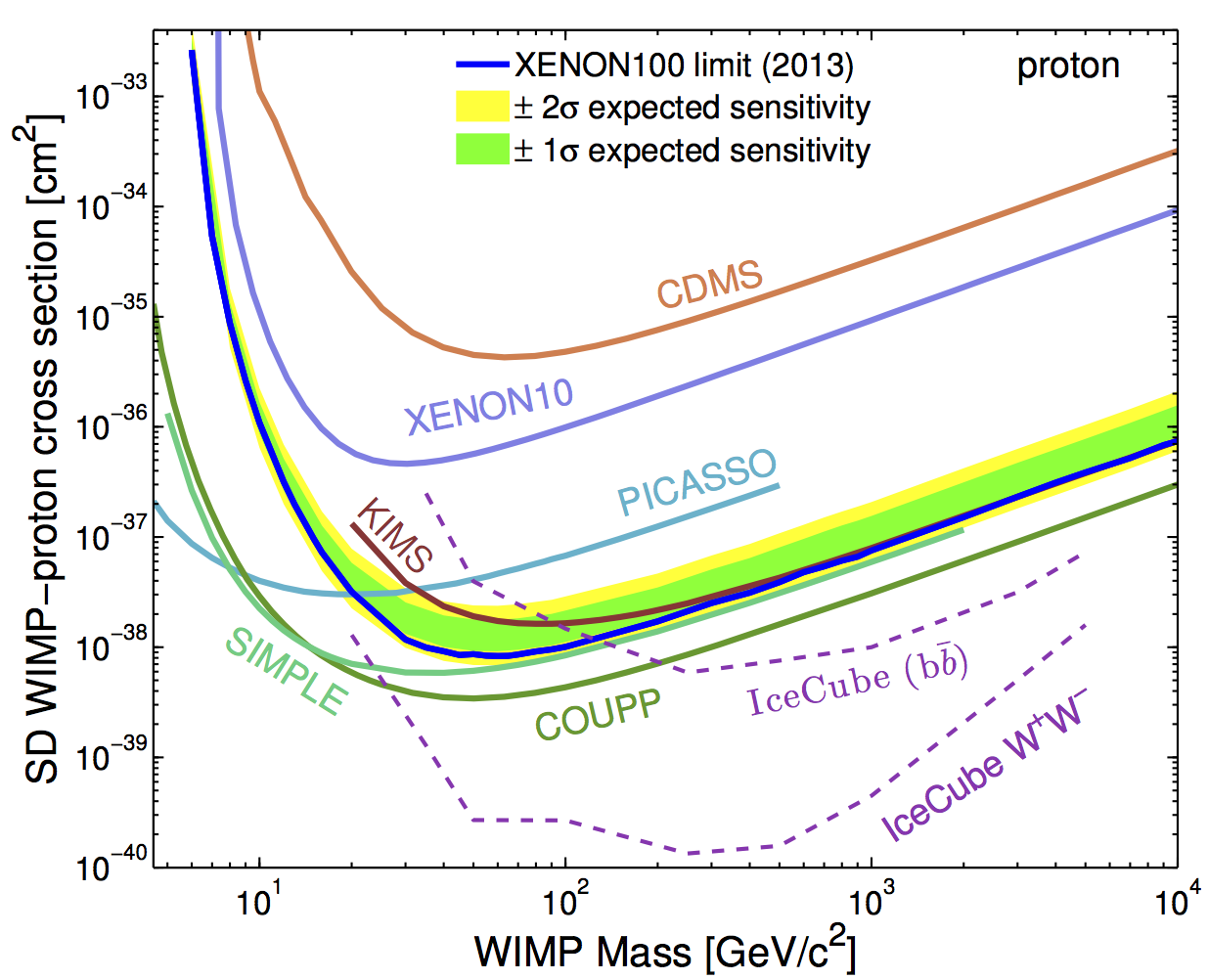}}
\end{minipage}
\caption{XENON100 90\% CL upper limits on the WIMP spin-dependent cross section on neutrons (left) and protons (right).}
\label{fig:xerun10sd}
\end{figure}

WIMPs in the halo of our Galaxy can interact in terms of scalar (or spin-independent, shown in Sec.\ref{subsec:run10si}) and axial-vector (or spin-dependent) couplings. If the WIMP is a spin-1/2 or a spin-1 field, the contributions to the WIMP-nucleus scattering cross section arise from couplings of the WIMP field to the quark axial current and will couple to the total angular momentum of a nucleus and only from nuclei with an odd number of protons or/and neutrons: in XENON100, the isotopic abundances of $^{129}$Xe (spin-1/2) and $^{131}$Xe (spin-3/2) are 26.2\% and 21.8\%, respectively. \\
The spin dependent differential WIMP-nucleus cross section is a proportional of the axial-vector structure function $S_A(q)$,\cite{sdxs} which we took from the large-scale shell-model calculations by Menendez et al.,\cite{mendez} that uses state-of-the-art valence shell interactions and less severe truncations of the valence space. For the first time chiral effective field theory (EFT) currents to determine the couplings of WIMPs to nucleons is also included in the calculation.\cite{eft} This yield a far superior agreement between calculated and measured spectra of the Xe nuclei, both in energy and in the ordering of the nuclear levels, compared to older results.\cite{russdea,tov} \\
Constraints on the spin-dependent WIMP-nucleon cross sections are calculated using the Profile Likelihood approach described and the same analysis selection of the spin-independent analysis.\cite{xe100run10si} XENON100 was able to exclude WIMP-neutron cross section down to $3.5 \times 10^{-40}$ cm$^2$ at a WIMP mass of 45 GeV/c$^2$ and set the most stringent limits to date on spin-dependent WIMP-neutron couplings for WIMP masses above 6 GeV/c$^2$, see Fig.\ref{fig:xerun10sd}.\cite{xe100run10sd}

\section{Nuclear recoil detector response}

\begin{figure}[h!]
\begin{minipage}{0.5\linewidth}
\centerline{\includegraphics[width=1.0\linewidth]{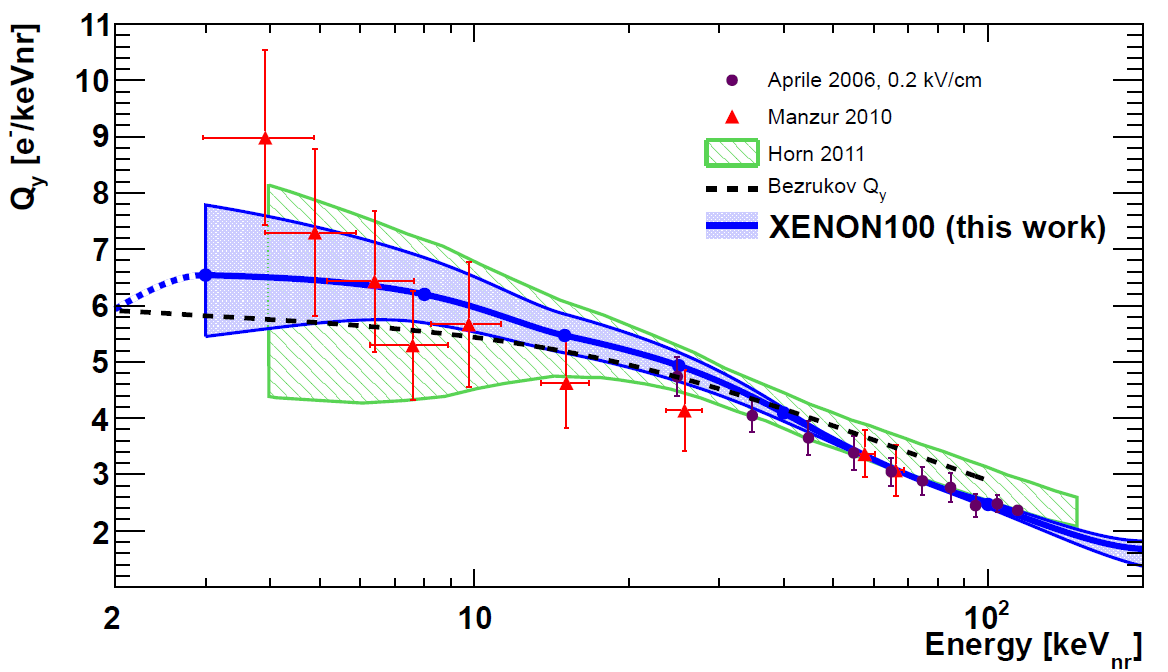}}
\end{minipage}
\begin{minipage}{0.5\linewidth}
\centerline{\includegraphics[width=1.0\linewidth]{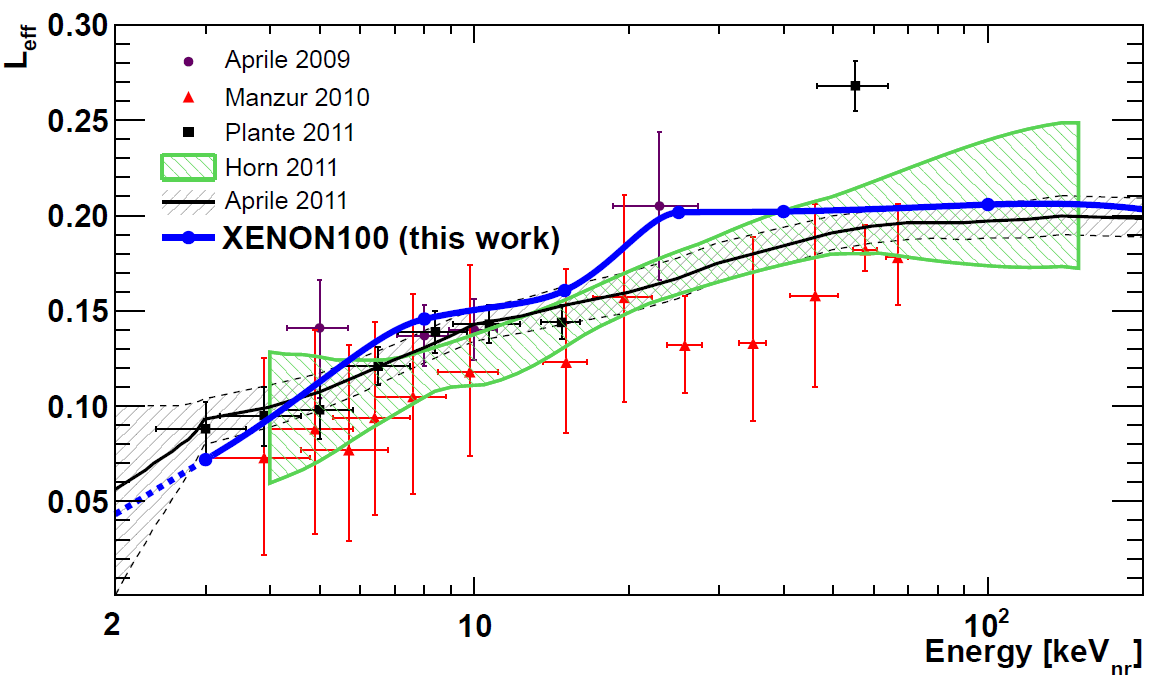}}
\end{minipage}
\caption{Left: Result on $\mathcal{Q}_{y}$  obtained from fitting the MC generated $S2$ spectrum to data. Right: $\mathcal{L}_{eff}$ obtained after optimization of the absolute $S1$ matching.}
\label{fig:xe100nresp}
\end{figure}

Commonly in double phase TCP data analysis the prompt scintillation ($S1$) and the proportional scintillation ($S2$) signals are treated independently. However, exploiting both channels together in allows considerably more robust constraints and better understanding of the detector response once data-MC matching is achieved, and both \leff ~and \qy ~can be extracted simultaneously. \\
We modeled the XENON100 detector response to $^{241}$AmBe calibration data by means of a MC that includes the signal generation in both the $S1$ and $S2$ channels. Agreement in the ionization channel is achieved through the adoption of a \qy ~that is largely consistent with previous direct and indirect measurements and phenomenological estimations but shows no indication of a low-energy increase as reported by the direct measurement, Fig.\ref{fig:xe100nresp} left. Additionally, an optimized \leff ~is determined using a similar method and is used to match data and MC signal distributions in the scintillation channel, Fig.\ref{fig:xe100nresp} right. Once the ionization and scintillation channels are combined in two-dimensional spaces it is possible to reproduce the data on the discrimination phase space which shows an excellent agreement (within 2\%) both means and widths of energy distributions with the data.\cite{xe100neuresp} \\
Using the same \leff ~employed in the spin-independent analysis and the \qy ~of Fig.\ref{fig:xe100nresp} left, we predicted the distribution of WIMP recoil events in the discrimination parameter space under the same configuration of the analysis described in Sec.\ref{subsec:run10si}. The excess of predicted WIMP recoil rate above only 2 event candidates observed in the 224.6 live-days XENON100 dark matter search is consistent with the reported exclusion limit, supporting the tension between these results and signal claims by other experiments.\cite{aal,ang,bern}

\section{XENON1T}

\begin{figure}
\centerline{\includegraphics[width=.67\linewidth]{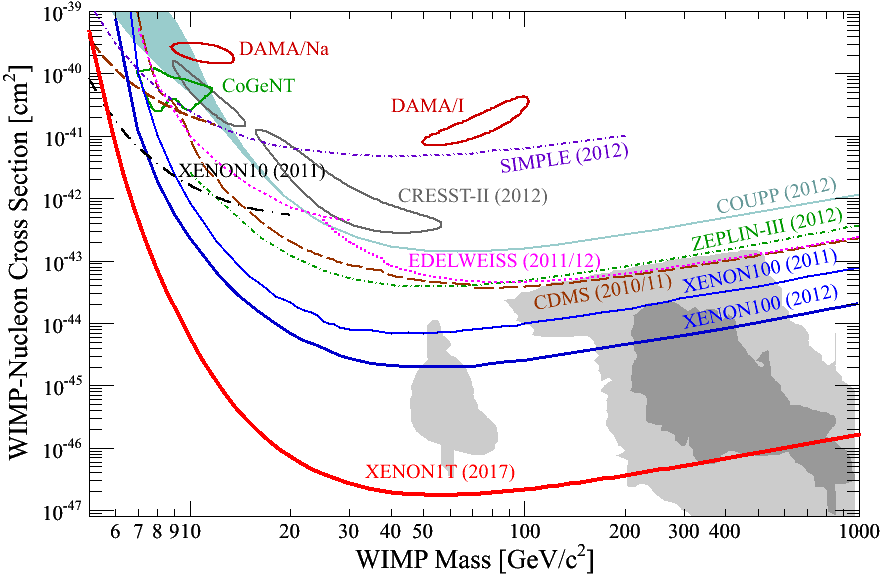}}
\caption{Achieved and projected limits on spin independent WIMP-nucleon cross-sections from the XENON100 and XENON1T detectors. For comparison, results from a selection of other experiments are shown along with the most likely parameter space for a detection as predicted by the Constrained Minimal Supersymmetric Extension of the Standard Model.}
\label{fig:xe1t}
\end{figure}

The next generation of the XENON project is represented by XENON1T, which will be a dual-phase TPC containing 3 tons of pure LXe with an cylindrical active volume of 1 m diameter and 1 m height. The site for the experiment has already been secured in Hall B at LNGS and the project has been fully funded. Technical designs for the water shield, service building, commissioning plan, and safety assessment are approved by the laboratory and the whole detector design is about finalized. We project the start of science data taking in 2015. Extensive material screening and state of the art LXe purification technology will enable a factor of 100 lower background with respect to XEONON100, reaching the level $5 \times 10^{-5}$ DRU which corresponds to less than one background event in two years of data taking in the energy range between 8 and 45 \kevnr. XENON1T will be able to probe WIMP interaction with spin-independent cross sections down to $2 \times 10^{-7}$ cm$^2$ for a mass of 50 GeV/c$^2$, Fig.\ref{fig:xe1t}, testing a wide range of Supersymmetric Extension of the Standard Model,\cite{susy} or will detect on the order of 100 dark matter events, for $10^{-45}$ cm$^2$ spin-independent cross section at 100 GeV/c$^2$.

%%%%%%%%%%%%%%


\section*{References}

\begin{thebibliography}{99}

\bibitem{dm} N. Jarosik et al., Astrophys. J. Suppl. 192, 14 (2011); K. Nakamura et al. (PDG), J. Phys. G37, 075021 (2010).

\bibitem{xe100instr} E. Aprile et al. (XENON100), \Journal\ASTR{35}{573}{2012}.

\bibitem{xe100run08} E. Aprile et al. (XENON100), \Journal\PRL{107}{131302}{2011}.

\bibitem{xe100pl} E. Aprile et al. (XENON100), \Journal\PRD{84}{052003}{2011}.

\bibitem{xe100run10si} E. Aprile et al. (XENON100), \Journal\PRL{109}{181301}{2012}.

\bibitem{sdxs} J. Engel, S. Pittel, and P. Vogel, {\em Int. J. Mod. Phys.} E1, 1 (1992).

\bibitem{mendez} J. Menendez, D. Gazit, and A. Schwenk, \Journal\PRD{86}{103511}{2012}.

\bibitem{eft} T. Park et al., \Journal\PRC{67}{055206}{2003}.

\bibitem{russdea} M. Ressell and D. Dean, \Journal\PRC{56}{535}{1997}.

\bibitem{tov} P. Toivanen et al.,  \Journal\PRC{79}{044302}{2009}.

\bibitem{xe100run10sd} E. Aprile et al. (XENON100), arXiv:1301.6620 (2013).

\bibitem{xe100neuresp} E. Aprile et al. (XENON100), arXiv:1304.1427 (2013). 

\bibitem{aal} C. Aalseth et al., \Journal\PRL{107}{141301}{2011}. 

\bibitem{ang} G. Angloher et al., \Journal\EPJ{72}{1971}{2012}. 

\bibitem{bern} R. Bernabei et al., \Journal\EPJ{67}{39}{2011} Eur. Phys. J. C67, 39 (2011).

\bibitem{susy} Combined region using C. Strege et al., JCAP 1203, 030 (2012); A. Fowlie et al., arXiv:1206.0264 (2012); O. Buchmueller et al., arXiv:1112.3564 (2011).

\end{thebibliography}
\end{document}